# Magnetic field tuning of terahertz Dirac plasmons in graphene

Hugen Yan,[1,§]* Zhiqiang Li,[2,§] Xuesong Li,[1] Wenjuan Zhu,[1] Phaedon Avouris[1]*

and Fengnian Xia[1]*

[1]IBM T. J. Watson Research Center, Yorktown Heights, NY 10598, USA
[2]National High Magnetic Field Laboratory, Tallahassee, Florida 32310, USA
[§] These authors contributed equally
*To whom correspondence should be addressed:
Email: hyan@us.ibm.com (H. Y.)
avouris@us.ibm.com (P. A.), fxia@us.ibm.com (F. X.)

(Dated: April 6, 2012)

**Abstract:**

**Boundaries and edges of a two dimensional system lower its symmetry and are usually regarded, from the point of view of charge transport, as imperfections. Here we present a first study of the behavior of graphene plasmons[1-6] in a strong magnetic field that provides a different perspective. We show that the plasmon resonance in micron size graphene disks[4,6] in a strong magnetic field splits into edge and bulk plasmon modes[7-12] with opposite dispersion relations, and that the edge plasmons at terahertz frequencies develop increasingly longer lifetimes with increasing magnetic field, in spite of potentially more defects close to the graphene edges. This unintuitive behavior is attributed to increasing quasi-one dimensional field-induced confinement and the resulting suppression of the back-scattering[13]. Due to the linear band structure of graphene[14,15], the splitting rate of the edge and bulk modes develops a strong doping dependence, which differs from the behavior of traditional semiconductor two-dimensional electron gas (2DEG) systems[7,8,11,16]. We also observe the appearance of a higher order mode indicating an anharmonic confinement potential[17] even in these well-defined circular disks. Our work not only**

**opens an avenue for studying the physics of graphene edges, but also supports the great potential of graphene for tunable terahertz magneto-optical devices.**

**Main text:**

Plasmon excitations in graphene are currently attracting great attention [1-6,12]. This is primarily due to the high carrier mobility and the tunable carrier density, which make graphene a promising plasmonic material in the far infrared and terahertz frequency ranges [1,2,5,6,18]. Unlike metal plasmons [19], the graphene plasmon is expected to be strongly affected by an external magnetic field due to a comparable cyclotron frequency [20-22] and plasmon frequency. The effects of a perpendicular magnetic field on the plasmons in conventional 2DEG systems have been extensively studied [7-11,17,23,24]. A major focus of these studies has been localized plasmons in disks [7] and submicron size quantum dots [8,17,23].

Here we report the first study of magneto-plasmons in micron size disk arrays of graphene- a different kind two dimensional system. We find that the plasmon lifetime can be dramatically modified by a magnetic field. This is especially true for the edge mode, which develops a much longer lifetime than that at zero field. This behavior is counterintuitive since imperfections of the edges may introduce more scattering and decrease the plasmon lifetime. To the best of our knowledge, this is the first demonstration of magnetic field tuning of plasmon lifetime in the terahertz frequency range.

Large area graphene disk arrays on $SiO_2$/Si are fabricated using standard electron beam lithography and dry etching techniques (for details, see Methods)[6]. Fig. 1a shows a scanning electron micrograph of a sample with a 3-micron diameter ($d = 3$ μm) disk array arranged in a triangular lattice, where the lattice constant is 4.5 μm. Fig. 1b shows the extinction spectrum ($1-T/T_S$) of such an array (Fermi level $E_f$= -0.54eV, determined from the cyclotron mass, see Supplementary Information) at zero magnetic field. The spectrum is referenced to the transmission $T_S$ through a bare substrate without graphene [18]. The dominant feature in this spectrum is a dipolar plasmonic resonance at ~130 cm$^{-1}$. As in the case of conventional 2DEG [7], this mode can be well described as a damped oscillator in the quasi-static limit as shown by the fitting curve (see Supplementary Information) [4,7]. However, in the high frequency side of the main peak in Fig. 1(b), there is an obvious deviation from the fit. A new mode emerges, whose origin will be discussed upon examination of the plasmon behavior in a magnetic field.

Fig. 2a shows the relative transmission spectra ($T(B)/T(0)$, referenced to the zero field transmission $T(0)$) of the graphene disk array shown in Fig. 1a in different magnetic fields up to 17.5 Tesla. These spectra can also be plotted as extinction spectra ($1-T(B)/T_s$), with reference to the bare substrate, as shown in Fig. 2b. The first order dipolar plasmon resonance splits into two modes, $\omega^+$ and $\omega^-$. Here, the line-widths change dramatically: the upper ($\omega^+$) branch broadens while the lower ($\omega^-$) branch narrows with increasing $B$-field. At the same time, the resonance of the higher order mode shifts up in frequency, but no splitting is observed. The inset of Fig. 2b shows the enlarged higher order mode portions for two spectra (at 14 and 17.5 T). The spectrum at 14 T is also fitted using a damped oscillator model (blue dashed line) and the difference between the measurement and fitting is obvious, clearly indicating the existence of the higher order mode. The oscillator strength of the higher order mode for single layer graphene is small. However, it appears more clearly for samples with two graphene layers due to stronger signals (see Supplementary Information).

The Fermi level in our graphene samples is usually greater than 0.3 eV (hole doping) [18]. When the magnetic field is below 20 Tesla, the measurement conditions remain in the quasi-classical regime [20], as is demonstrated by the linear magnetic field dependence of the cyclotron resonance frequency for the unpatterned graphene (see Supplementary Information) [20,21]. Assuming that carriers with mass $m_c$ in the disk are confined by a parabolic potential, the real part of the diagonal dynamic conductivity in a magnetic field for the array is given by (see Supplementary Information):

$$\mathrm{Re}\,\sigma_{xx}(\omega, B) = \frac{fD}{2\pi}\left[\frac{\omega^2\Gamma}{(\omega_0^2 + \omega\omega_c - \omega^2)^2 + \omega^2\Gamma^2} + \frac{\omega^2\Gamma}{(\omega_0^2 - \omega\omega_c - \omega^2)^2 + \omega^2\Gamma^2}\right], \qquad (1)$$

where $f$ is the filling factor (percentage of the graphene coverage on the sample) of the array, $D$ is the Drude weight [18], $\Gamma$ is the scattering rate, $\omega_0$ is the plasmon frequency at zero magnetic field, and $\omega_c = eB/m_c$ is the cyclotron resonance frequency. In the quasi-classical regime, the cyclotron mass is $m_c = |E_f|/v_f^2$, where $E_f$ is the Fermi energy and $v_f$ is the Fermi velocity [20,25]. In a traditional 2DEG the approximation of a parabolic confinement potential of the charge carriers in the disk usually describes the first order dipole resonance in a magnetic field very well [11]. However, the appearance of higher order modes in the spectrum is not accounted for, and additional modification of the confinement potential is needed [17,23]. In a magnetic field, analysis of equation(1) reveals that the original plasmonic resonance splits into two modes with frequencies:

$$\omega^{\pm} = \sqrt{\omega_0^2 + \omega_c^2/4} \pm \omega_c/2 \,. \quad (2)$$

Fig. 2c shows the extracted peak frequencies obtained by fitting the spectra using two independent damped oscillators [2](see Supplementary Information) as a function of the magnetic field. The frequency difference between the upper and lower branches, which is the cyclotron frequency according to equation(2), is also shown, and the higher order mode frequency is plotted as well. The $\omega^+$ and $\omega^-$ branches, and their difference, are fitted using equation(2), with fitting parameters $\omega_0 = 134 cm^{-1}$ and $m_c = 0.096 m_0$, where $m_0$ is the free electron mass.

Fig. 2d shows the extracted full width at half maximum (FWHM) $\Delta\omega$ for the $\omega^+$ and $\omega^-$ branches. The lower branch plasmon lifetime ($\tau = 1/\Delta\omega$) increases from 79 femto-seconds to 132 femto-seconds at 17.5 Tesla, while that for the upper branch decreases to 47 femto-seconds. These trends can be fitted using a relation for the FWHM derived directly from equation(1):

$$\Delta\omega^{\pm} = \Gamma \pm \frac{2\omega_c \Gamma}{\sqrt{4\omega_0^2 + (\omega_c+\Gamma)^2} + \sqrt{4\omega_0^2 + (\omega_c-\Gamma)^2}} \,, \quad (3)$$

where the cyclotron mass is the same as used in Fig. 2c and the fitting parameter $\Gamma = 76$ cm$^{-1}$.

Equation(3) shows that the upper branch line-width can approach 2Γ, while that for the lower branch decreases as $1/B^2$ in the high magnetic field limit. This fitting is reasonably good and the minor deviation may be due to the doping inhomogeneity.

The plasmon lifetime tuning in the magnetic field as evidenced by the observed resonance line-width change is remarkable, especially the sharpening of the lower frequency mode. Fig. 3a shows a spectrum of a less doped graphene disk array ($E_f$= -0.34 eV) where the effect is even more pronounced. The bulk mode line-width is 5 times larger than that of the edge mode. The predicted line-width of the edge mode using equation(3) is only 20 cm$^{-1}$, which corresponds to a lifetime of 250 femto-seconds. The measured line-width for the spectrum in Fig. 3a is about 30 cm$^{-1}$, larger than the predicted value, probably due to the doping inhomogeneity. Typically, there are two factors contributing to the plasmon line-width. One is the radiative plasmon decay[19], in which the plasmon is coupled to the electromagnetic radiation field. The radiative decay can be the dominant contribution to the plasmon resonance line-width in noble metal particles. The other originates from the scattering of participating carriers in the collective motion by impurities and defects[19]. Based on the first mechanism, the plasmon lifetime can be tuned by radiative engineering of metal particle arrays[26]. In the case of the graphene plasmon in the terahertz frequency range discussed here, the radiative decay has minimal effect on the lifetime since the line-width of the plasmon resonance corresponds very well to the Drude scattering rate (see Supplementary Information). As a result, the magnetic field dependence of the plasmon line-width indicates that the field can suppress (for the $\omega^-$ branch) or enhance (for the $\omega^+$ branch) the carrier scattering.

In order to understand this effect, we have to consider the different nature of the two modes. In a magnetic field, the plasmon in a bulk two dimensional system has an energy gap $\hbar\omega_c$ (the cyclotron energy), which implies that the plasmon frequency should be larger than the single particle cyclotron frequency[10,11]. The higher frequency mode ($\omega^+$ branch) observed here has this character, and corresponds to the bulk mode. As illustrated in Fig. 3b, at high magnetic field, the carriers participating in the bulk mode undergo cyclotron motion collectively inside the disk and are hardly affected by the confinement potential of the disk. Eventually, in the high field limit ($\omega_c \gg \omega_0$), this mode becomes the cyclotron resonance, with the line-width doubled compared to the zero field plasmon resonance (For cyclotron resonance line-width, see Supplementary Information) [20].

Appearance of the plasmon mode with lower frequency ($\omega^-$ branch) than the plasmon energy gap $\hbar\omega_c$ is attributed to the edge effect, since at the edge, the Landau level bending leads to metallicity and the absence of a gap [10,11]. This mode is the so-called edge magneto-plasmon [9]. It features a current along the edge and hence a rotating dipole with frequency $\omega^-$. The width of the edge is on the order of the cyclotron radius [10,11] which is 30 nm at 17.5 Tesla for the sample in Fig. 2. Fig. 3b also illustrates the skipping orbits of the participating carriers [13]. At the edge, a scattering event by defects or impurities cannot reverse the current in a strong magnetic field, as illustrated by the orbits in Fig. 3(b) [13]. It is the suppression of the back-scattering that is responsible for the long lifetime of the edge magneto-plasmon. As in the quantum Hall effect [13], edge effects play a crucial role in the behavior of this plasmon mode.

Interestingly, unlike in graphene, the line-width for the split plasmonic peaks in traditional 2DEG disks does not seem to follow equation(3) in the terahertz frequency range [7,8,17]: the line-widths for both modes ($\omega^+$ and $\omega^-$) were found to be similar and to slightly decrease with increasing magnetic field. However, measurements in the radio frequency regime (Megahertz to Gigahertz) did show sharp edge plasmon resonances [9].

In addition, compared to traditional 2DEG, the cyclotron mass depends on the Fermi energy for Dirac Fermions in graphene[14,15], as expected from the expression $m_c = |E_f|/v_f^2$. As a consequence, the splitting rate of the two plasmon branches can be tuned in graphene by doping, either electrostatically or chemically [18]. Fig. 4 shows the cyclotron frequencies for three samples with different doping levels, determined from the splitting of the upper and lower branches in the magnetic field. The doping levels are controlled by baking or nitric acid exposure time durations(see Methods). The extracted cyclotron masses are plotted in the inset of Fig. 4 as a function of the zero field plasmon frequencies. Since the plasmon frequency is proportional to $\sqrt{E_f}$ [1,2], the cyclotron mass versus plasmon frequency has a parabolic scaling relation, as shown by the solid line in the inset of Fig. 4.

Finally, we return to the discussion of the weak higher order mode. This mode is not described by equation(1) derived for a parabolic confinement potential. Indeed, the generalized Kohn's theorem [24,27] in the dipole approximation does not allow for the appearance of the higher order mode in the absorption spectrum for carriers confined by a parabolic potential. In our case, the typical wavelength is one order of magnitude larger than the graphene disk diameter, and the dipole approximation is satisfied. Appearance of the higher order mode is therefore an indication of the breakdown of Kohn's theorem[23]. In slightly doped graphene under a strong magnetic field, this theorem is inherently inapplicable because of the linear band structure of graphene [28,29]. However, at higher doping levels, the Landau levels near the Fermi surface are not well separated, and Kohn's theorem is asymptotically valid [30]. Therefore, we attribute the breakdown of Kohn's theorem to the anharmonicity of the confinement potential. Indeed, in previous studies of traditional 2DEG quantum dot with the shape of the dots deviating from the circular shape, higher order edge modes were observed [8,11,23], obviously due to the anharmonic confinement introduced by the dot geometry. However, the higher order mode observed in our graphene disks does not evolve into an edge mode in a magnetic field, which is different from the first order mode as well as other higher order modes observed previously in the traditional 2DEG quantum dots [8,23]. This mode is a second order dipolar mode which was previously predicted by theory [17]. The inset of Fig. 1B depicts the charge distribution for this mode and the first order dipolar mode. To the best of our knowledge, this is the first time that this mode has been observed experimentally. Calculation of the frequency of this mode in a magnetic field according to theoretical predictions [17] (see Supplementary Information) results in excellent agreement with the experimental observations (Fig. 2c).

In summary, we show that the plasmon lifetime in graphene can be continuously tuned using a magnetic field and that the mode splitting rate can be tuned through doping. In addition, the observation of a higher order dipolar mode indicates the anharmonicity of the confinement potential in graphene disks. The results demonstrate that graphene is a very promising material for plasmonics and magneto-optics studies, and that magneto-plasmon spectroscopy opens a new avenue to study graphene edge effects. Other intriguing phenomena may be found in the quantum regime, where the carriers are distributed in only a few Landau levels.

During the preparation of this manuscript, we became aware of a related work [31] presented by Crassee *et al*. at APS March Meeting. In that work, magneto-plasmons were observed in unpatterned epitaxial graphene on SiC due to the inherent domain boundaries.

**Methods**

In this work, single layer graphene grown on copper foil through chemical vapor deposition is used [32]. The graphene on copper is first transferred to a highly resistive silicon substrate with a 90 nm thermal oxide layer. The graphene disk arrays are then fabricated using standard e-beam lithography and plasma etching. The disks are arranged in a triangular lattice, and the total area for a typical array is 3.6 mm by 3.6 mm. The as-prepared graphene samples are hole-doped with Fermi energy of around 0.3 eV. If exposed to nitric acid, the doping level can be enhanced dramatically [6,18]. On the other hand, baking the samples at 170ºC in air usually leads to reduction in doping concentration.

During the measurement, the samples and reference silicon substrates are mounted in a sample holder, which is inserted into a liquid helium cryostat and cooled via helium exchange gas to a temperature of 4.5 Kelvin. Unpolarized infrared light from a Bruker IFS-113 Fourier transform spectrometer is delivered to the sample via a light pipe, and the transmitted light is detected by a composite Si bolometer. The measurements are performed in a superconducting magnet in the Faraday geometry (magnetic field perpendicular to the sample surface). The transmission through the graphene disk array $T$ and through the reference substrate $T_s$ are measured in a magnetic field from 0 to 17.5 Tesla.

**Figure captions**

**Figure 1**: **Scanning electron micrograph and far-infrared characterization of the graphene disk array at a zero magnetic field.**
**a**, A scanning electron micrograph of a graphene disk array. The diameter of the disk ($d$) is 3 microns and the lattice constant ($a$) is 4.5 microns. **b,** An extinction spectrum at a zero magnetic field. The solid curve is a fitting according to the plasmon resonance model (damped oscillator) described in the Supplementary Information. The deviation in the high frequency side of the peak is due to the higher order mode. The charge distributions of the first and second order dipolar modes are illustrated in the inset.

**Figure 2**: **Magnetic field tuning of the plasmon resonance.**
**a,** Transmission $T(B)$ through the disk array at field $B$, referenced to the zero field transmission. The second order dipolar mode is indicated. **b,** Extinction spectra of (**a**) with a zero field spectrum also shown. The inset is a zoom-in for two of the spectra to emphasize the higher order mode. The spectrum measured at 14 T is also fitted using the damped oscillator model (dashed blue line). **c**, Magnetic field dependence of the peak frequencies, where the difference frequency for $\omega^+$ and $\omega^-$ is also shown. Solid curves for first order modes are fits based on equation(2). The solid curve for the second order mode is a calculated result described in the Supplementary Information. **d**, Magnetic field dependence of the FWHM of the plasmon resonances. Solid curves are fits using equation(3).

**Figure 3**: **Edge and bulk modes.**
**a**, An extinction spectrum for a less doped sample at 16 Tesla. The solid curve is a fit based on damped oscillators (Supplementary Information). **b**, Illustration of the charged carrier motion for the edge and bulk modes. The blue dots represent scatterers.

**Figure 4**: **Doping dependence of the mode splitting.**
The splitting as a function of magnetic field for three samples with different doping levels. Solid lines are linear fits. The inset shows the extracted cyclotron mass for these three samples as a function of the zero field plasmon resonance frequency. The solid curve represents the parabolic scaling.

**Acknowledgements:**
The authors are grateful to B. Ek and J. Bucchignano for technical supports, to T. Low and V. Perebeinos for helpful discussions, and to D. Farmer for proof reading of this manuscript. Part of this work was performed at the National High Magnetic Field Laboratory, which is supported by NSF/DMR-0654118, the State of Florida, and DOE.

**Author contributions:**
H. Y., Z. L., P. A. and F. X. conceived the experiments. H. Y. and F. X. fabricated the devices, with assistance from W. Z.. Z. L. performed the measurements. X. L. provided graphene on copper foils. H. Y. analyzed the data. H. Y. and F. X. co-wrote the paper. P. A. supervised the project and all authors commented on the paper.

Figure 1

a

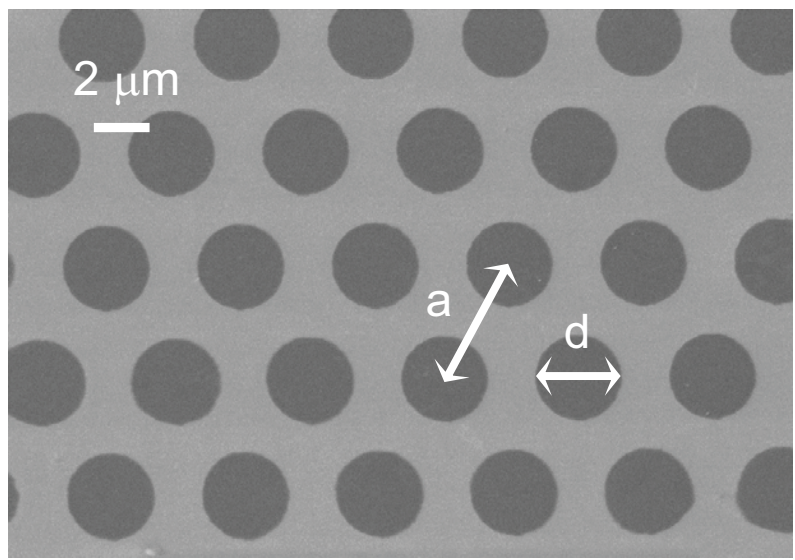

b

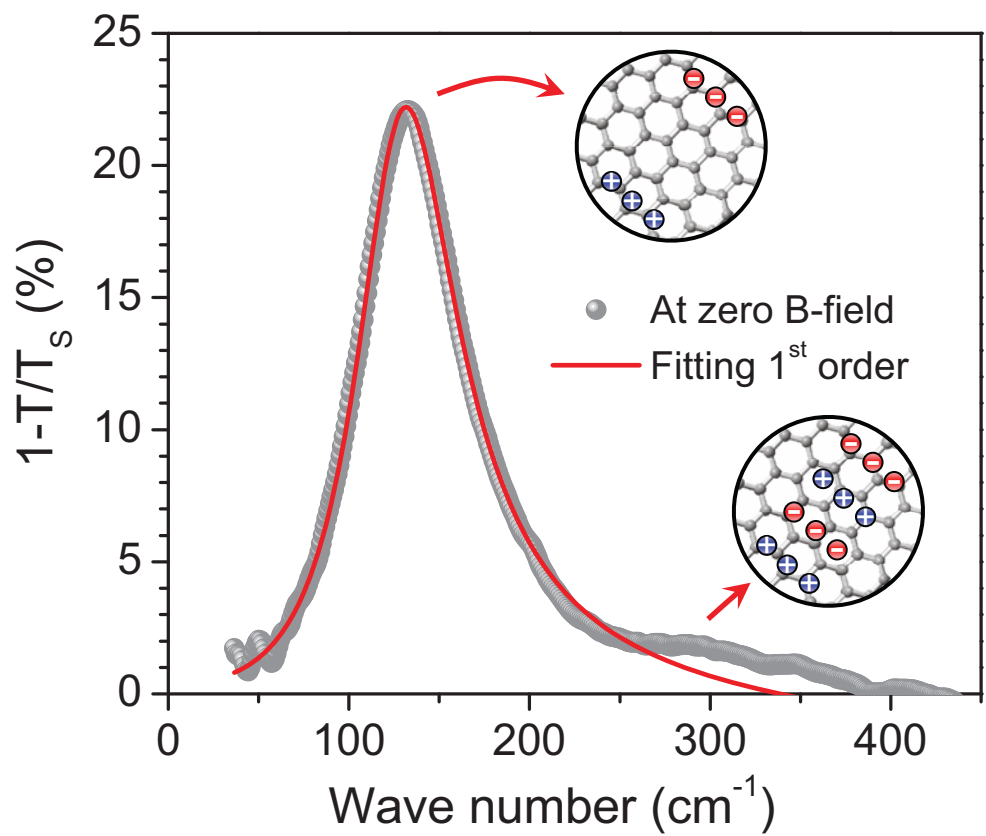

Figure 2

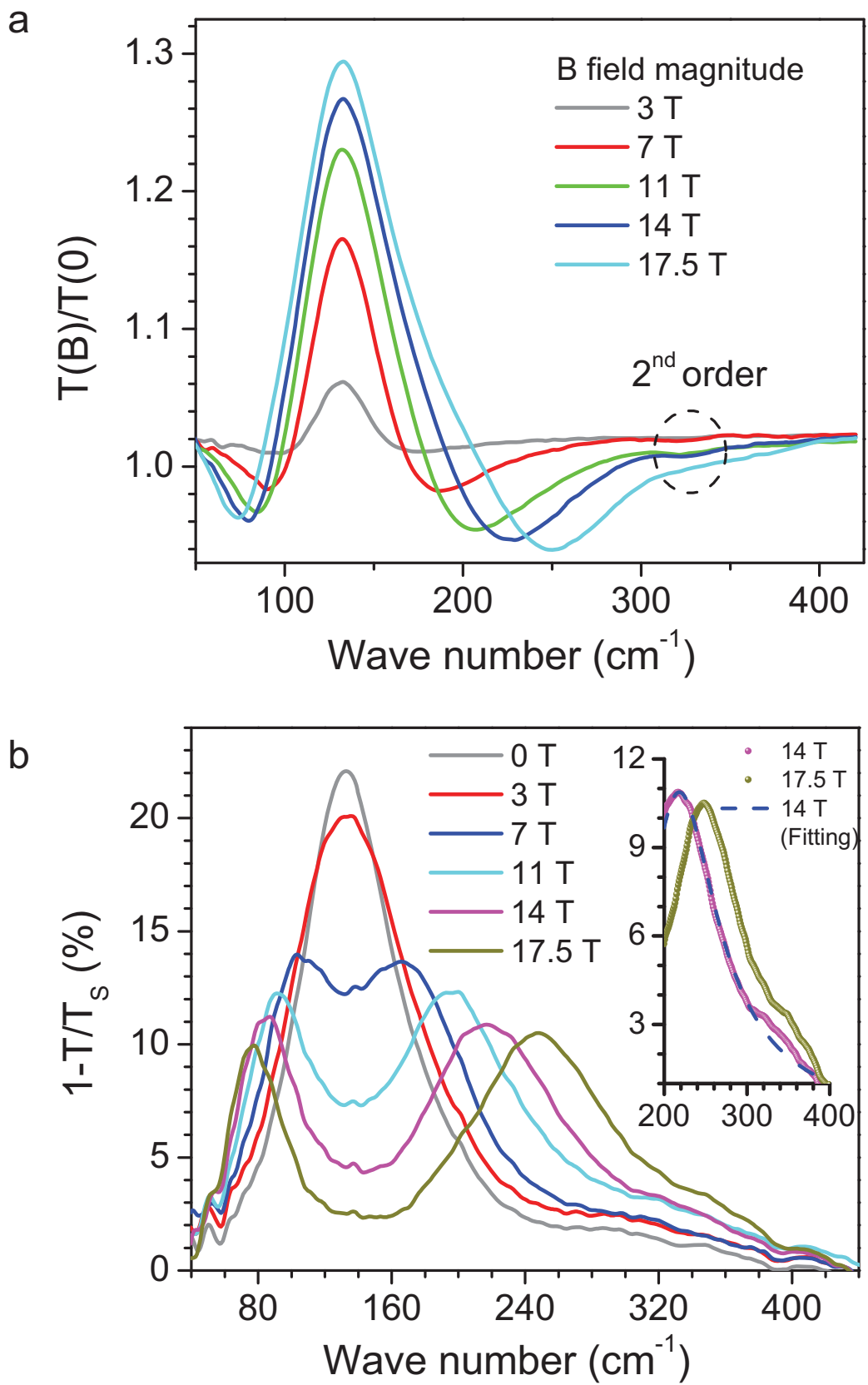

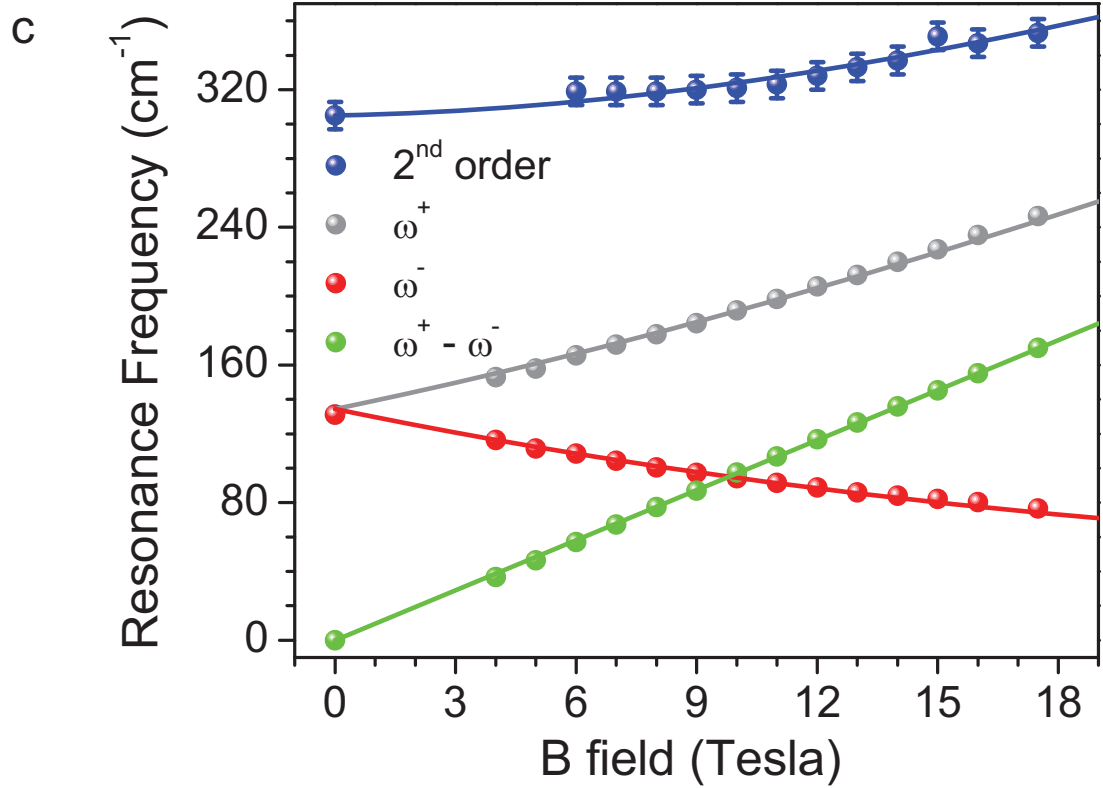
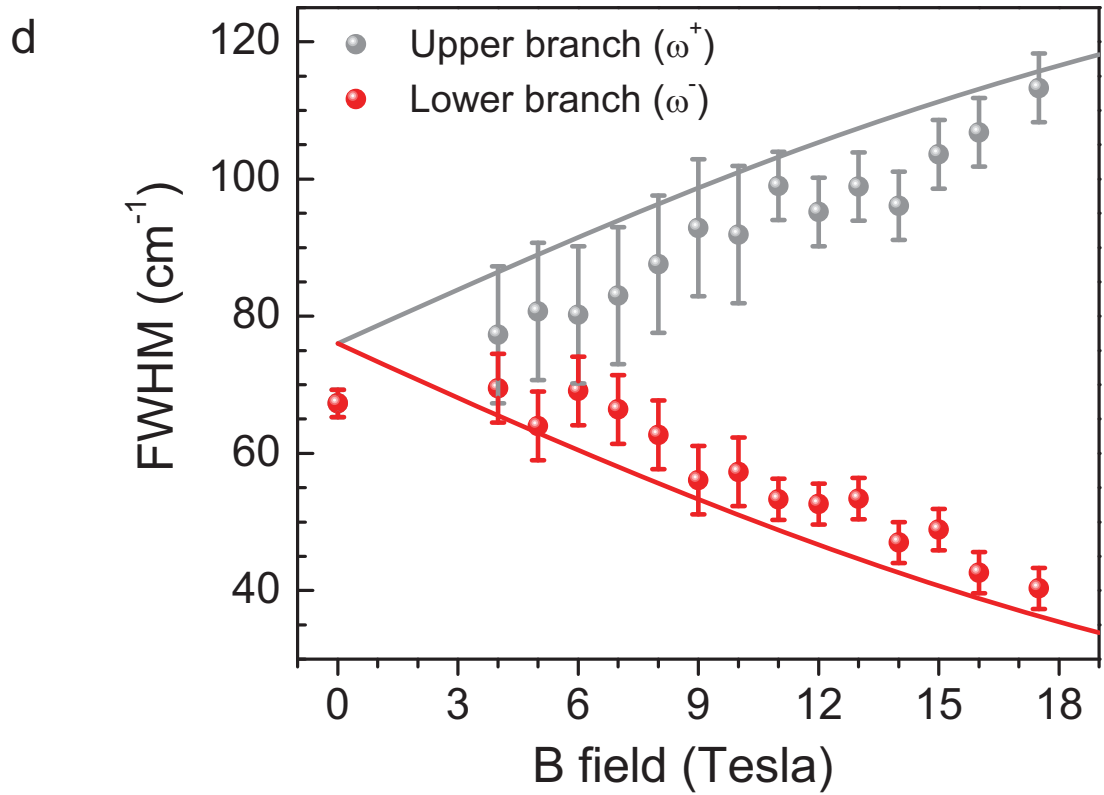

Figure 3

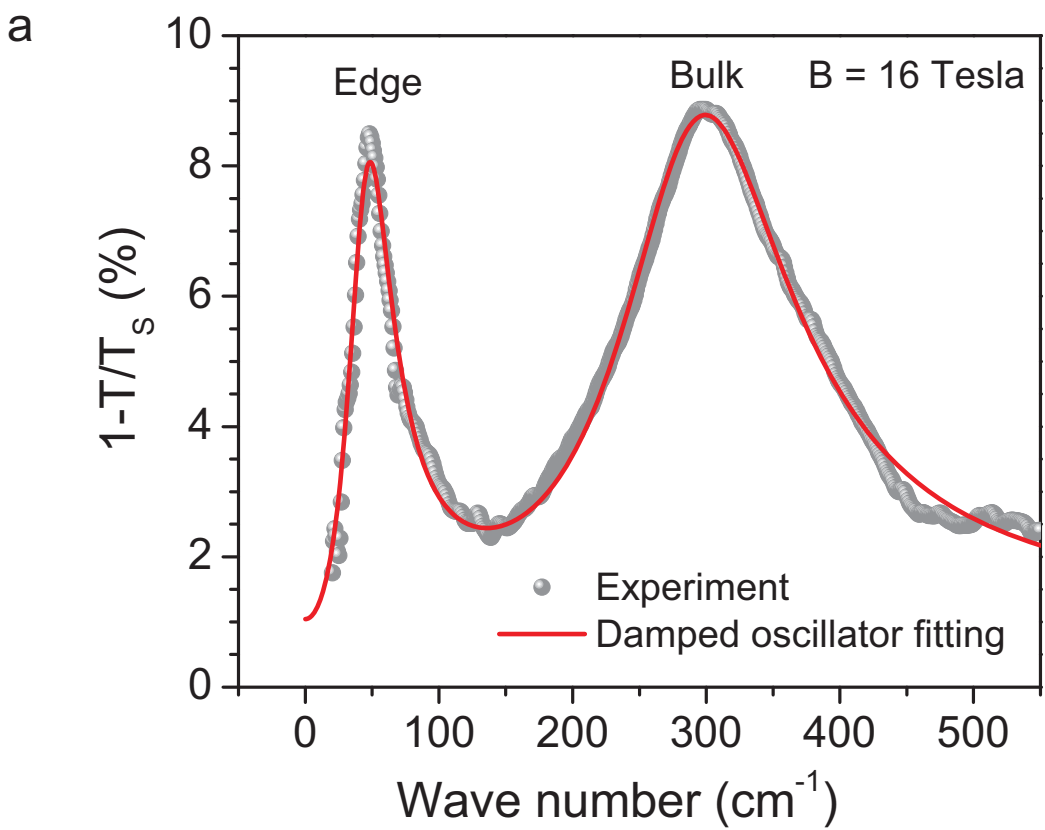

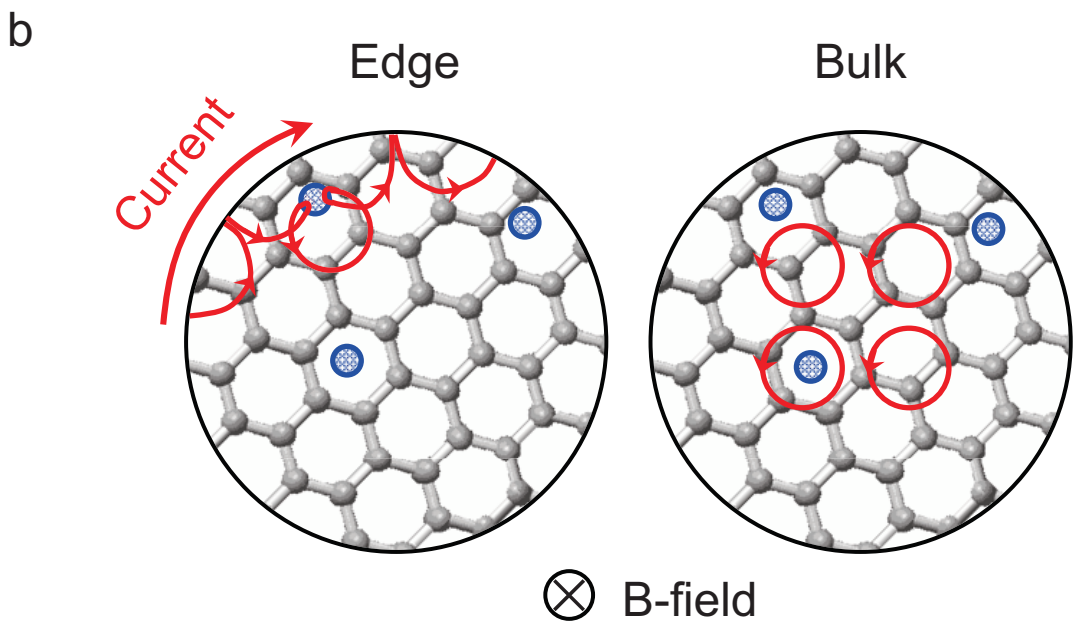

Figure 4

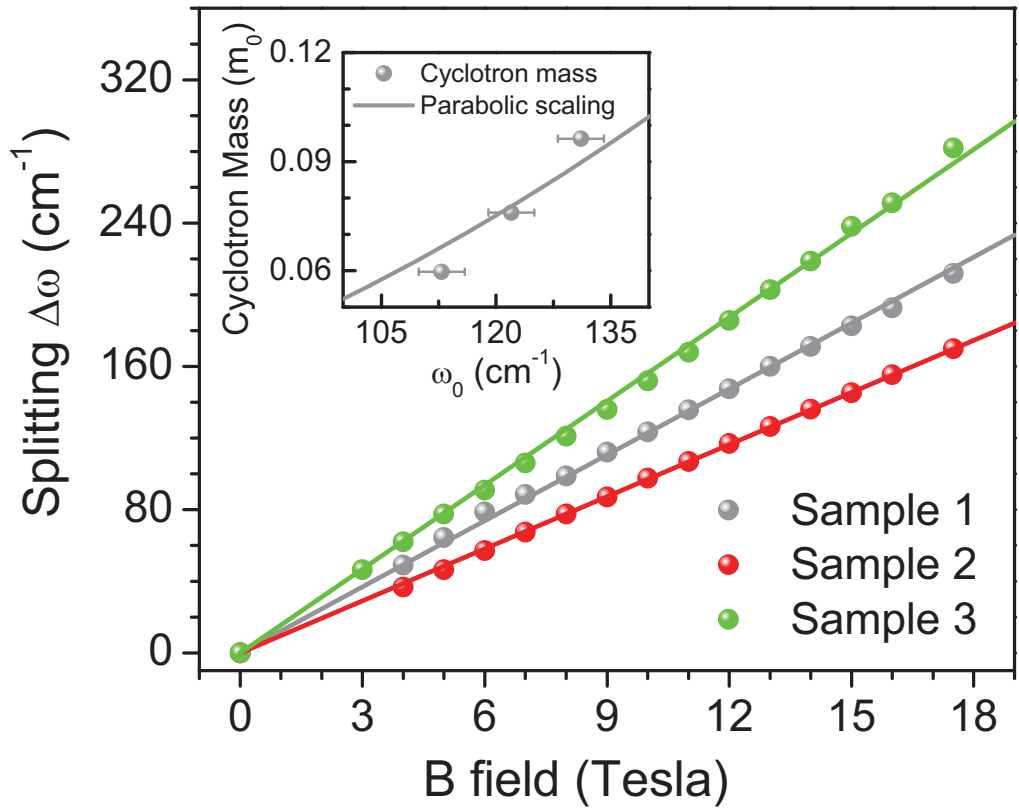

Supplementary Information for

# Magnetic field tuning of terahertz Dirac plasmons in graphene

Hugen Yan,* Zhiqiang Li, Wenjuan Zhu, Phaedon Avouris*
and Fengnian Xia*

To whom correspondence should be addressed:
email: hyan@us.ibm.com (H. Y.), avouris@us.ibm.com (P. A. ), fxia@us.ibm.com (F. X.)

**Contents:**

1. **Extinction and the dynamic conductivity**
2. Cyclotron resonance of the unpatterned graphene
3. Plasmon resonance and its line-width
4. Derivation of equation (1) in the main text
5. Calculation of the second order dipolar mode frequencies
6. Additional data for a two-layer disk array

## 1. Extinction spectra and the dynamic conductivity

For graphene on a substrate with index of refraction $n_s$, the light extinction $1-T/T_s$ is related to the complex dynamic conductivity $\sigma(\omega)=\sigma'(\omega)+i\sigma''(\omega)$ [S1,2]:

$$1-T/T_s = 1 - \frac{1}{\left|1+Z_0\sigma(\omega)/(1+n_s)\right|^2}, \tag{S1}$$

where $Z_0$ is the vacuum impedance $\sqrt{\mu_0/\varepsilon_0}$ and $\omega$ is the frequency of the light. When the extinction is small, it only depends on the real part of the conductivity $\sigma'$ and equation (S1) can be simplified to:

$$1-T/T_s = \frac{2Z_0\sigma'(\omega)}{1+n_s} \tag{S2}$$

Therefore, the dynamic conductivity can be determined by the measurements of the extinction spectra.

## 2. Cyclotron resonance[S3] of the unpatterned graphene

The extinction spectra of the unpatterned graphene sheet in a magnetic field were also measured. Fig. S1a shows the spectra at different fields up to 17.5 Tesla. We plot the extracted resonance frequencies and the full widths at half maximum (FWHM) in Fig. S1b. The frequency has a linear dependence on the magnetic field ($\omega_c = eB/m_c$), as indicated by the linear fitting with a cyclotron mass $m_c = 0.071m_0$. The FWHM is about 150 cm$^{-1}$ for all of the fields, which is 2 times of the Drude scattering rate $\Gamma$ determined from the Drude response at zero magnetic field ($\Gamma = 75$ cm$^{-1}$ for this sample and all other samples exhibit similar values). The linear dependence of the cyclotron resonance[S3] observed in our samples indicates that the measurements were performed in a quasi-classical regime [S4]. This is different from the $\sqrt{B}$ dependence of the cyclotron frequency observed in the quantum regime, where Landau levels are well-resolved due to a lower Fermi energy [S5,6].

The FWHM of the cyclotron resonance at different magnetic fields remains almost constant, which is a classical result and can be described by the Drude model. In the Drude framework, the diagonal dynamic conductivity of the graphene sheet in a magnetic field is [S3,4,7]

$$\sigma_{xx}(\omega,B) = \frac{iD}{\pi}\frac{\omega+i\Gamma}{(\omega+i\Gamma)^2 - \omega_c^2} \tag{S3}$$

where $D$ is the Drude weight, $\Gamma$ is the Drude scattering rate, and $\omega_c$ is the cyclotron frequency. equation (S3) indicates that the FWHM is $2\Gamma$ and does not vary with the magnetic field.

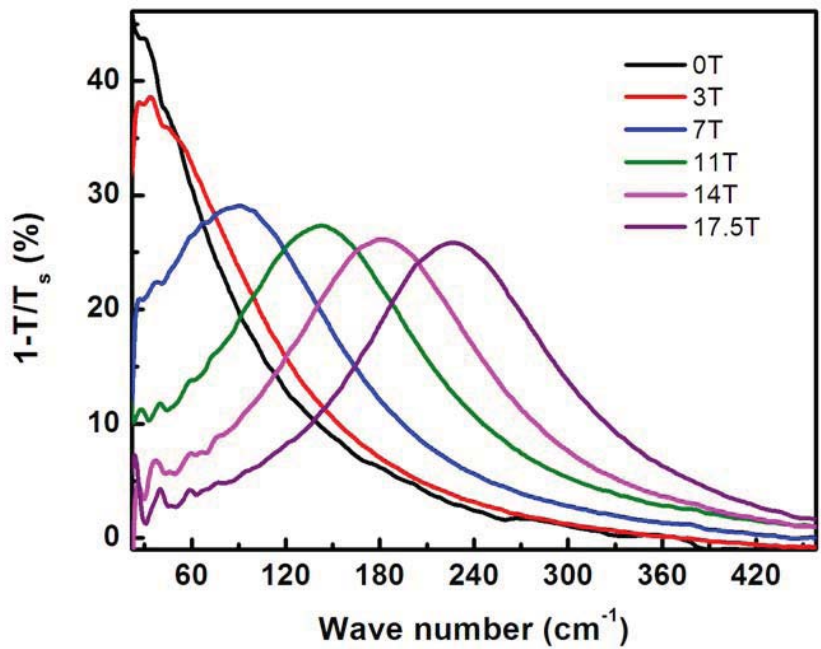

a

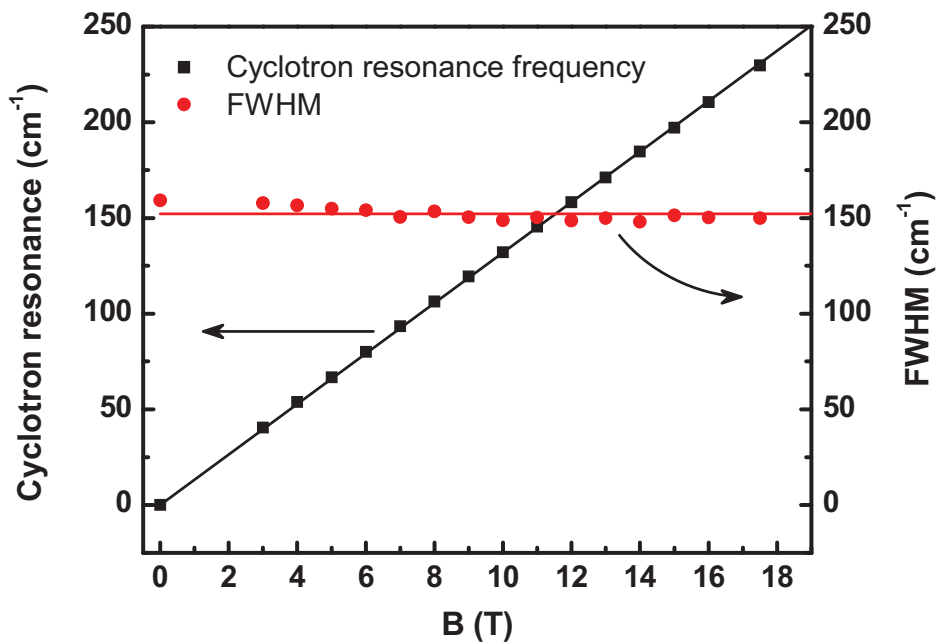

b

**Figure S1**. **Cyclotron resonance of an unpatterned graphene sheet. a**, The extinction spectra in different magnetic fields. **b**, The extracted resonance frequencies and FWHM. The resonance frequency depends on the *B*-field linearly and the horizontal line is a guide to the eye to show the constant FWHM.

## 3. Plasmon resonance and its line-width

The plasmon resonance of a graphene disk array has been derived previously based on the polarizability of an oblate spheroid [S8]. The dynamic conductivity at zero magnetic field for an array with filling factor $f$ (the percentage of graphene disk coverage) is [S8]:

$$\sigma(\omega) = i\frac{fD}{\pi}\frac{\omega}{(\omega^2 - \omega_0^2) + i\Gamma\omega}, \quad (S4)$$

where $\omega_0$ is the plasmon resonance frequency. equation S4 above has the form of a damped oscillator. The fitting curves in Fig. 1b and Fig. 3a of the main text are based on equation S2 and the real part of equation S4. For the fitting in Fig. 3a, two such independent damped oscillators are used. Equation S4 implies that the FWHM of the plasmon resonance is $\Gamma$, half of that for the cyclotron resonance. Fig. S2 shows a plasmon resonance spectrum for a graphene disk array with $d$ of 2.2 μm along with a cyclotron spectrum at 17.5 T. The FWHM for the plasmon resonance is ~70±10 cm$^{-1}$, about half of that of the cyclotron resonance. Since the line-width of the plasmon resonance is almost identical to that of the Drude scattering rate, the carrier scattering dominates the plasmon line-width in this case, and other mechanisms such as radiative decay are negligible.

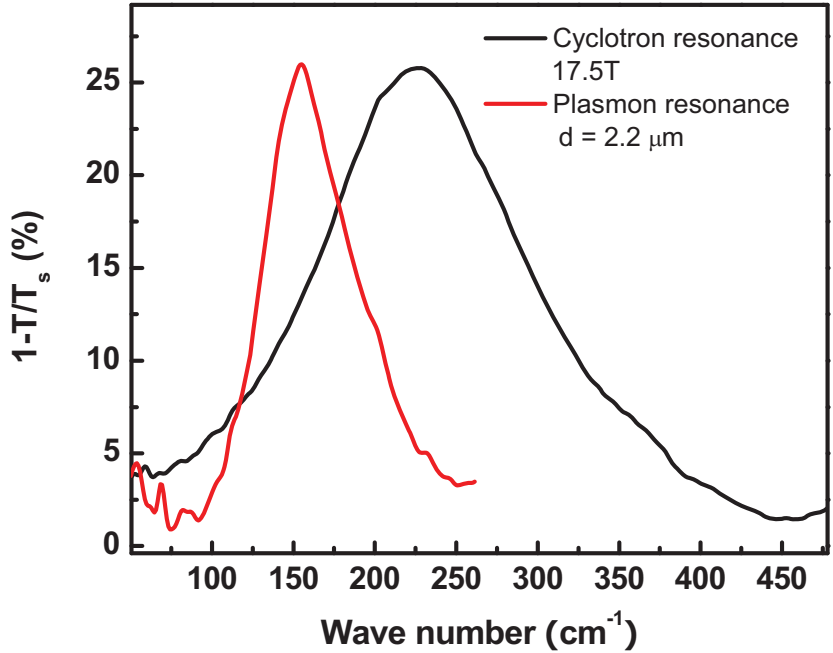

**Figure S2. Comparison of cyclotron and plasmon resonance spectra.** The plasmon spectrum is normalized to have the same peak height as that for the cyclotron spectrum.

From a quantum mechanical perspective, cyclotron resonance involves the transition between Landau levels. Since each of the initial and final energy levels has broadening Γ, the cyclotron resonance line-width is 2Γ. On the other hand, the plasmon resonance only involves one energy level with broadening Γ and the electro-magnetic radiation creates or annihilates plasmon quanta on that level. As a result, the plasmon resonance has a line-width of Γ.

## 4. Derivation of equation (1) in the main text

The dynamic conductivity (equation (1) in the main text) for a graphene disk array can be modeled using a harmonic oscillator in a magnetic field. We treat Dirac fermions classically and the mass is $m_c = |E_f|/v_f^2$. The confinement potential within the disk is assumed to be parabolic (harmonic):

$$V(x,y) = \frac{1}{2} m_c \omega_0^2 (x^2 + y^2), \tag{S5}$$

where $\omega_0$ is the plasmon frequency at zero magnetic field. According to the generalized Kohn's theorem [S9-11], the collective excitation of $N$ charged particles with mass $m_c$ and charge $e$ confined by a parabolic potential has the same infrared extinction spectrum as that of a single particle with mass $Nm_c$ and charge $Ne$. Therefore, the dynamic conductivity can be calculated based on a single particle theory. The equation of motion for a single confined particle in a magnetic field perturbed by a small infrared electrical field $E(t) = E_0 e^{-i\omega t}$ along the $x$ direction reads:

$$\begin{aligned} m_c \ddot{x} + m_c \Gamma \dot{x} + m_c \omega_0^2 x &= eE_0 e^{-i\omega t} + e\dot{y}B \\ m_c \ddot{y} + m_c \Gamma \dot{y} + m_c \omega_0^2 y &= -e\dot{x}B \end{aligned}, \tag{S6}$$

where Γ is the scattering rate. The dynamic conductivity (equation (1) in the main text) is obtained by solving the equation of motion above.

## 5. Calculation of the second order dipolar mode frequencies

The appearance of the higher order mode is due to the anharmonicity of the confinement potential. Interestingly, accurate numerical simulations show that the parabolic confinement formalism can still be used to estimate the magnetic field dependence of the resonance frequency of the second order dipolar mode [S12], as long as the measured resonance frequency ($\omega_{30}$) at zero $B$-field is used in such calculations. This conclusion also applies to the determination of the resonance frequency of the first order modes. The equation governing the frequency dispersion of the second order dipolar mode is [S12]:

$$\omega^3 - (\omega_c^2 + \omega_{30}^2)\omega - \frac{1}{11} \omega_c \omega_{30}^2 = 0 \tag{S7}$$

where $\omega_{30}$ is the measured resonance frequency at zero $B$-field. The solid curve for the second order mode in Fig. 2c of the main text is calculated with $\omega_{30} = 305 \text{cm}^{-1}$, cyclotron mass $m_c = 0.096 m_0$ (the same as used for the first order modes).

## 6. Additional data for a two-layer graphene disk array

Measurements performed on disk arrays consisting of multiple graphene layers usually have better signal to noise ratio [S13]. Fig. S3 shows additional magneto-plasmon data for a disk array (diameter $d$ = 5 μm, lattice constant $a$ = 7.5 μm) with two vertically stacked graphene layers with similar doping concentration. This figure is the two-layer counterpart of the single layer data shown in Fig. 2 of the main text. With stronger absorption, the second order dipolar mode is more pronounced, as shown in the inset of Fig. S3a. In general the behavior of the two-layer graphene disk array in a magnetic field is similar to that of the single layer counterpart.

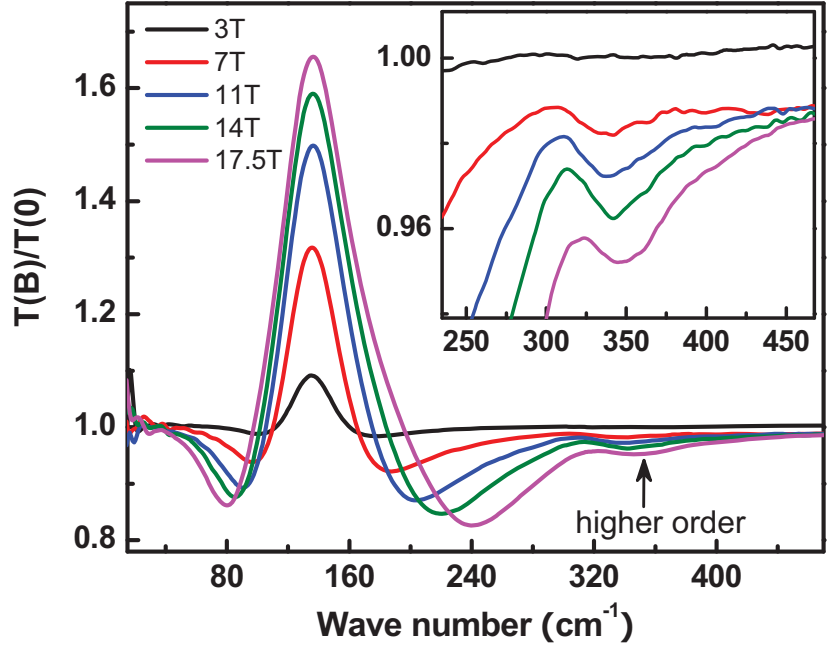

a

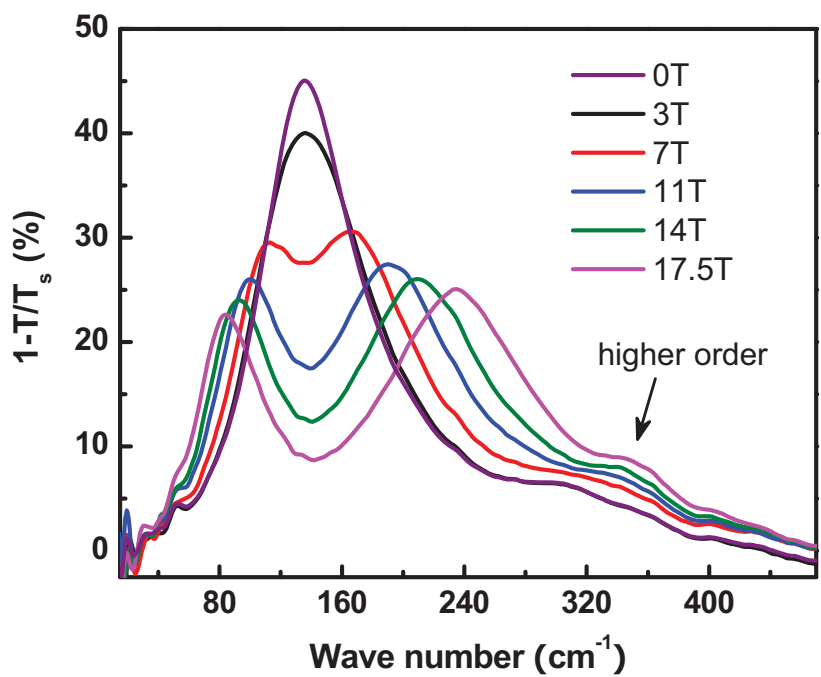

b

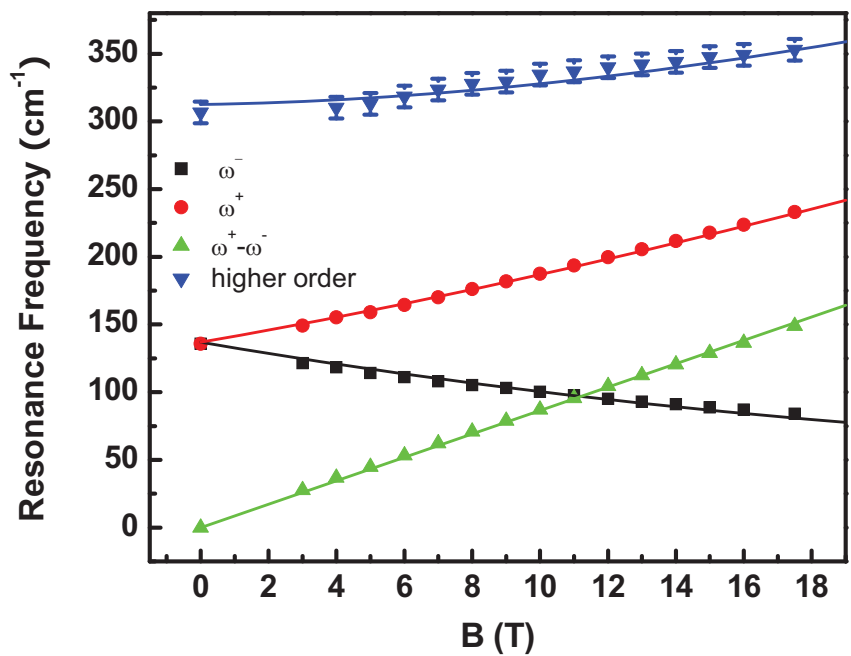

c

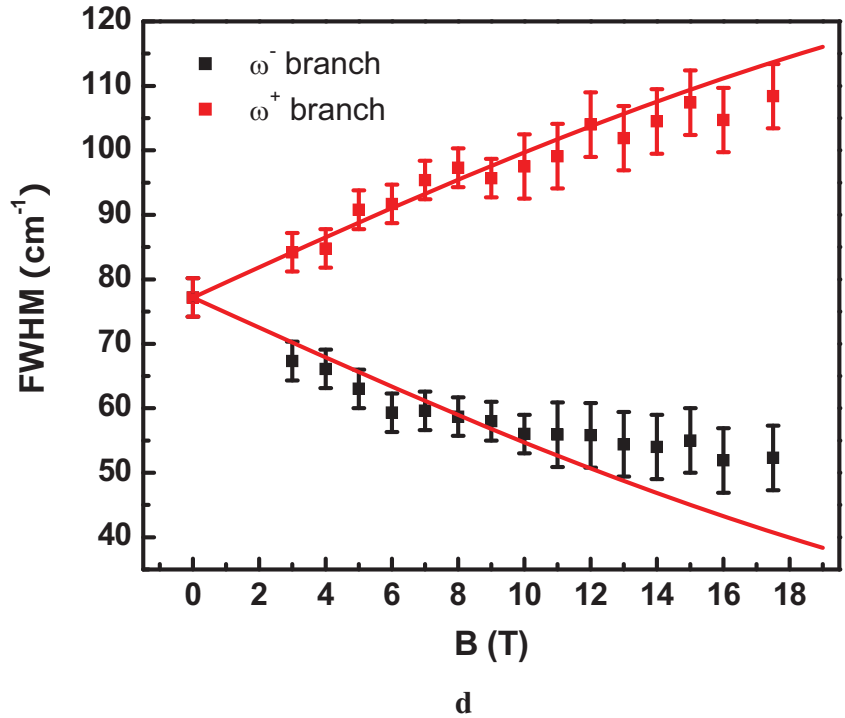

**Figure S3**. **The magneto-plamon resonance data of a disk array consisting of two vertically stacked graphene layers**. **a**, Transmission $T(B)$ through the disk array at field $B$, referenced to the zero field transmission. The second order dipolar mode is indicated. The inset shows the enlarged portion of the second order mode. **b**, Extinction spectra of (a) with a zero field spectrum also shown. **c**, Magnetic field dependence of the peak frequencies, where the difference frequency for $\omega^+$ and $\omega^-$ is also shown. Solid curves for first order modes are fits based on equation (2) in the main text. The solid curve for the second order mode is a calculated result described in equation S7. **d**, Magnetic field dependence of the FWHM of the plasmon resonances. Solid curves are fits according to equation (3) in the main text.